\documentclass[aps,pre,notitlepage,superscriptaddress,floatfix]{revtex4-1}
\usepackage{graphicx}
\usepackage{color}
\usepackage{dcolumn}
\usepackage{bm}

\begin{document}
\preprint{APS/123-QED}
\title{Poiseuille flow of soft polycrystals in 2D rough channels}
\author{Tanmoy Sarkar}
\affiliation{Department of Physics,  Indian Institute of Technology, Bombay, Powai, Mumbai-400 076, India.}
\author{Pinaki Chaudhuri}
\affiliation{Institute of Mathematical Sciences, CIT Campus, Taramani, Chennai 600113, India  }
\author{Anirban Sain}
\email{asain@phy.iitb.ac.in}
\affiliation{Department of Physics,  Indian Institute of Technology, Bombay, Powai, Mumbai-400 076, India.}
\date{\today}
\begin{abstract}

Polycrystals are partially ordered solids where crystalline order extends over mesoscopic length scales, namely, the grain size.
We study the Poisuielle flow of such materials in a rough channel.
In general, similar to yield stress fluids, three distinct dynamical states, namely, flowing, stick-slip and jammed can be observed, 
with a yield threshold dependent on channel width. Importantly, the 
interplay between the finite channel width, and the intrinsic ordering scale (the grain size) leads to  
new type of spatiotemporal heterogeneity. In  wide channels, although the average flow profile remains plug like, 
at the underlying granular level, there is vigorous grain remodelling activity resulting from the velocity heterogeneity among the 
grains. As the channel width approaches typical grain size, the flowing polycrystalline 
state  breaks up into a spatially heterogeneous mixture of flowing liquid like patches and chunks of nearly static grains. Despite 
these static grains, the average velocity still shows a parabolic profile, dominated by the moving liquid like patches. However,
the solid-liquid front moves at nearly constant speed in the opposite direction of the external drive. 
\end{abstract}
\pacs{Valid PACS appear here} 
\maketitle

\section{Introduction}

Dense colloidal emulsions are good model systems for studying polycrystalline and amorphous 
solids. Polycrystals have mesoscopically large ordered regions, the grains, which are separated 
by atomistically narrow disordered grain boundaries, across which the grain orientation changes 
abruptly \cite{gokhale2013grain}. In contrast,  there is no spatial order in atomic arrangements, 
for amorphous materials. So, in terms of 
the degree of spatial heterogeneity, polycrystals fall in between single crystal and amorphous 
material \cite{zhang2018compression}. How does the intrinsic ordering scale, the grain size, make
polycrystals different in terms of dynamic response ? It has been shown that sheared polycrystals, 
share some of the flow properties (vortices, saddles etc.) of amorphous material 
\cite{gokhale2013grain,santi}, yet retaining unique features like grain rotation, dislocation creep etc. 
Experimental studies involving rheology of soft polycrystals have started to probe these questions
\cite{gokhale2012directional,tamborini2014plasticity, buttinoni2017colloidal, Lavergne6922}.

Driven flow of such materials through a rough channel (e.g., under gravity or pressure gradient)
requires that the applied force overcome the resistive  friction from the channel walls.
But the wall friction, due to no slip boundary condition, gives rise to velocity gradients
and thereby can create a spatially non-uniform stress field.
Can the polycrystal sustain these stress gradients or does it break down into 
a flowing amorphous state, when it yields ?
In this Letter, we probe this question.



In this letter, using molecular dynamics (MD) simulations, we report the flow 
response of a 2D  soft polycrystalline system, confined in a rough channel 
and  subjected to a constant body force (e.g., gravity), resulting in a 
Poiseuille flow.  We demonstrate that the onset of steady flow depends 
on a yield threshold which decreases as the channel width is increased. 
Surprisingly, for wide channels, i.e., when $w \gg d$, the particle 
diameter, despite maintaining a polycrystalline morphology, a plug-like flow 
emerges, while grains continuously undergo break-up and coalescence within the plug. 
On the other hand, for  narrow channels, when $w \approx 10d$, there 
is significant melting of the polycrystalline structure and interestingly, the 
fluid-solid front moves at a constant speed in the opposite direction of the 
main flow.


\section{Model} 

For our study, we considered a weakly bi-disperse, two-dimensional Leanard-Jones (LJ) 
system with total number of particles $N=12118$. By choosing the fraction of large 
particle as $1\%$ and the size ratio of $1.4$, we generate polycrystalline states 
\cite{shiba2010plastic}. We worked at volume fraction $\phi=1$ and temperature $T=0.2$. 
The weakly bi-disperse system can sustain a polycrystalline non-equilibrium steady 
state, under shear \cite{Onuki, sarkar2017grain}, unlike the monodisperse system 
which tends to form single crystal when sheared. The 2D Poiseuille flow is set up 
in a rectangular box ($L\times w$) with two confining walls perpendicular to the 
y-axis  and periodic boundary conditions along x axis. We apply driving force to an 
initial state which is a preformed polycrystalline sample prepared using the above 
parameters. To generate the Poiseuille flow, only the particles within a slab, 
of width $w$, parallel to $\hat x$ axis, centred around the middle of the box, were 
subjected to an additional constant external force $F_{ext}\hat x$. The static 
particles above and below the slab, which interact via the LJ potential with 
the flowing particles, serve as the rough channel walls. The width $w$ of the 
slab was systematically varied to study the flow for different confinements. 
The temperature is maintained via a local thermostat, implemented by dividing 
the system into thin slabs parallel to the walls and rescaling the velocity 
regularly after every few iterations. 

\section{Results}

\begin{figure}[]
\includegraphics[width=7.5cm]{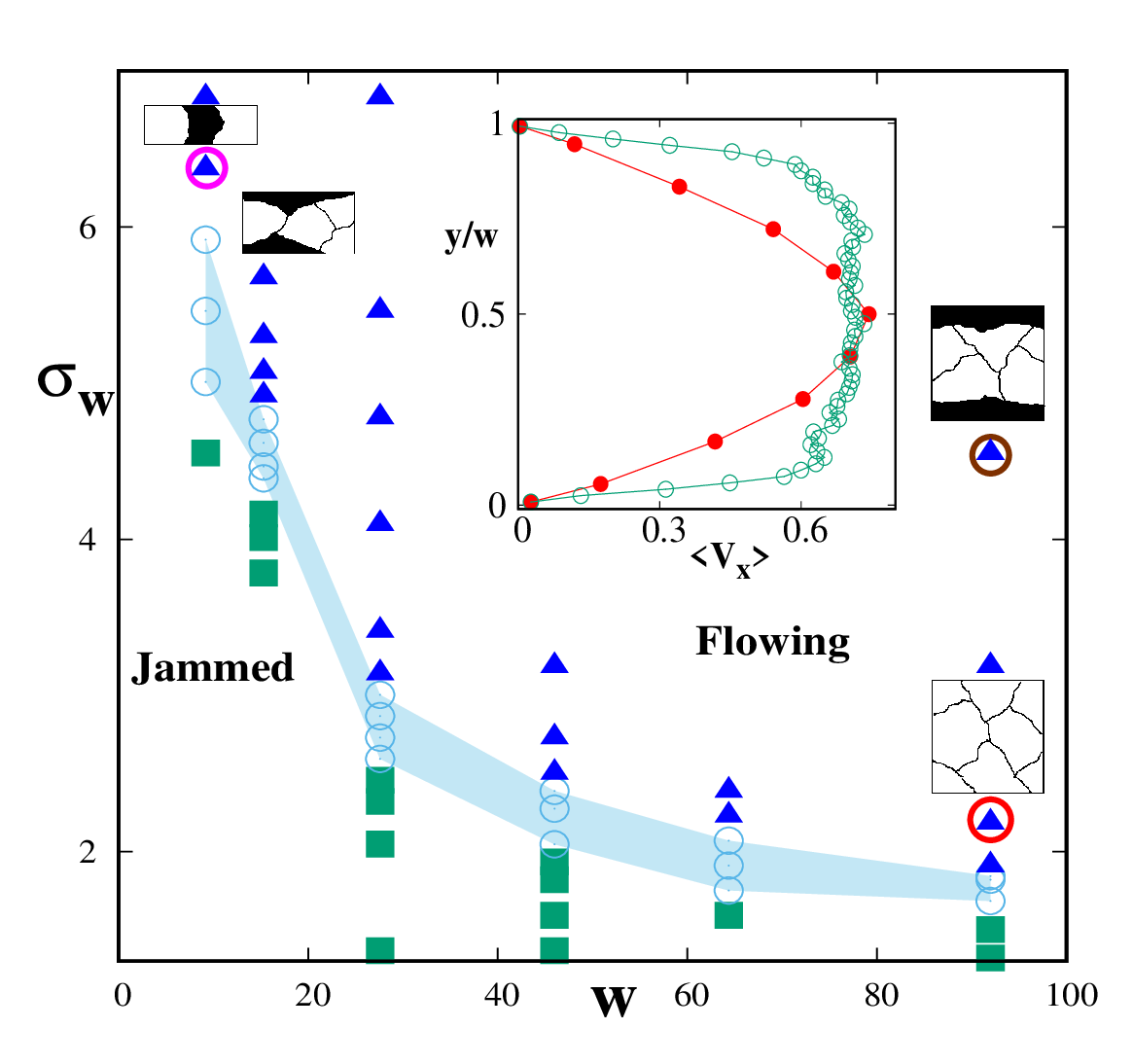}
\caption{Phase plot as a function of channel width $w$ and wall stress $\sigma_w$. Three different dynamical 
states are identified: flowing (triangles), stick-slip (circles) and jammed (squares). The shaded region represents 
the stick-slip states.  In the flowing state stratification of the system into polycrystalline and liquid like zones 
(shaded)  are indicated by schematic diagrams. The inset shows the averaged velocity profile $v_x(y)$ plotted 
as a function of $y/w$.  The flow exhibits  plug and parabolic profiles,  for wide (open circles) and narrow 
(solid circles) channels, respectively.
The corresponding parameters are ($w,\sigma_w)= (9.2,6.4)$ and $(92,2.2)$.}
  \label{fig.phase_dia}
\end{figure}

Our MD simulation shows that despite some similarities with amorphous
solids at macroscopic scale \cite{goyon, mansard2013molecular, 
chaudhuri2014poiseuille}, the flow behaviour of confined polycrystal 
is distinct. 
Although the externally applied body force ($F_{ext}$) is uniform, the interplay with the confining walls 
create spatially non-uniform stress.  The polycrystalline solid 
yields due to this applied forcing,  when the effective wall stress ($\sigma_w=\rho F_{ext} w/2$) 
exceeds a threshold  which depends upon the channel width $w$ 
\cite{chaudhuri2012dynamical}. Fig.\ref{fig.phase_dia} shows an 
approximate state diagram, describing the state of the system (flowing or
jammed) for a wide range of channel widths and driving forces.
Further, we observe that while unjamming with the increase in $F_{ext}$, an intermediate
state  with stick-slip motion is observed  (the shaded region in Fig.\ref{fig.phase_dia}); 
also see Fig-S1 in Supplementary Information (SI)\cite{epaps}. 
Such states are also observed in yielding of amorphous
systems \cite{varnik2008profile} and even in channel flow of
monodisperse colloids \cite{genovese2011crystallization}.  

In the inset of Fig.\ref{fig.phase_dia} we show average velocity 
profiles, for different channel widths ($w$), in steady state conditions.
For large  $w$, the flow profile of the soft polycrystal 
exhibits behaviour similar to yield stress fluids : $v_x(y)$ 
is plug-like at the central bulk region and has high, but 
nearly constant velocity gradient near the walls, resembling a boundary layer. 
In contrast, for narrow channels, the averaged flow profile 
$v_x(y)$ is parabolic, similar to low Reynolds number
(viscous) pipe flow. 
This is consistent with the observations in dense, soft, glassy material
\cite{goyon} where such tarnsition from plug to parabolic flow  with
decreasing channel width was rationalised via the existence of non-local
spatial correlations in plasticity during flow. 
Beyond such similarities with amorphous solids in the average
flow profiles, some distinct features arise in flowing polycrystals 
due to its underlying grain dynamics, and that is the focus of
this work.  In Fig.\ref{fig.phase_dia}, we show  schematic pictures
of the distinct grain morphologies (detailed later) that occur at 
different parts of the state diagram.

\begin{figure}
\centerline{ \includegraphics[width=8cm]{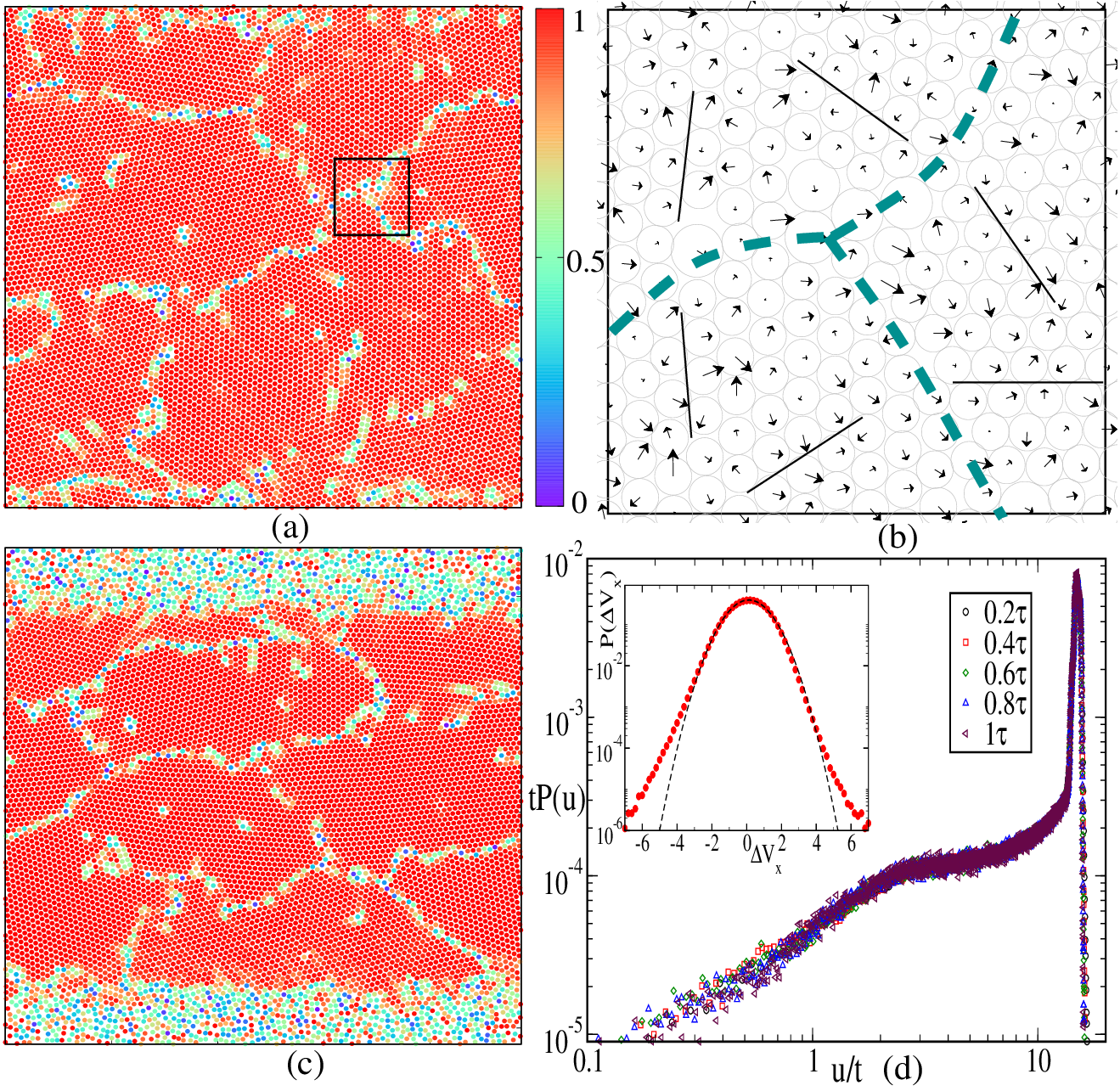}}
\caption{Flowing grains in a wide channel ($w=92$), at low (a) and high 
(c) driving forces, $\sigma_w=2.2$ and $4.6$, respectively. Colour bar 
shows hexatic order parameter, $\Psi_6$, value. A significant boundary layer (BL) emerges at high 
force. A triple grain junction in (a) is zoomed in subplot (b), and velocity 
vectors added, in order to show the velocity  heterogeneity among grains.
(d) Scaled displacement distributions for different time intervals $t$ 
(symbols shown in legend-box) collapse. The inset shows distribution of 
nongaussian velocity fluctuation (for case-c) along the flow.  
}
  \label{fig2}
\end{figure}

Now we discuss the typical steady state flow behaviour in the case of the 
wide channels, at low (Fig.\ref{fig2}-a) and high (Fig.\ref{fig2}-c)
forces (see the state points marked in Fig.\ref{fig.phase_dia}). 
While fluid-like structure prevails in the boundary layer (more prominent in 
Fig.\ref{fig2}-c than in Fig.\ref{fig2}-a), polycrystalline grain  structure is 
maintained in the bulk. 
The bulk, in both cases, exhibits plug-like flow profile,
with the plug front moving at higher velocity at higher force.
However, the grain structure, in both cases, constantly remodels 
via grain break-up and coalescence \cite{sarkar2017grain} (see movie 
in SI~\cite{epaps}). In Fig.\ref{fig2}(a,c), we use the $\Psi_6$ hexatic order 
parameter (OP) representation to capture the local arrangement 
of the particles and also to clearly demarcate the grain boundaries.

{This dynamic cycle of grain growth and breakage occurring in the bulk can be
attributed to  heterogeneity in the velocities of the particles across grains. 
The enclosed box in the bulk of Fig.\ref{fig2}(a) is zoomed in subplot 
Fig.\ref{fig2}(b) to illustrate this more clearly. It shows a moving triple junction of 
grains, where the grain boundaries are shown by broken lines. We see that 
particles constituting each grain have transverse motions in different directions
within each grain. Similar  velocity map for Fig.\ref{fig2}(c), at  higher force (and higher particle 
velocities), is shown and explained in the SI ({Fig.S4}) \cite{epaps}.}
Further quantitative analysis in terms of non-homogeneous transverse velocities and strain 
rates, causing slow shear of the grains, are given by Fig.S2, Fig.S3 in SI~\cite{epaps}.
We also illustrate that there are finite fluctuations in transverse velocity, 
$\langle\delta v^2_y(y)\rangle$, in the bulk, indicating existence of plastic events. 
Relatedly, there is indeed nonzero strain rate $\dot\gamma =\partial v_x/\partial y$ 
(shown in Fig-S3 of SI\cite{epaps}), at the centre of the channel.
However, the magnitude of the local shear-rate is much  weaker at the middle 
compared to the boundary regime. This local shear leads to the continuous  
grain growth and break up near the centre. This heterogeneous dynamics can be 
quantified by studying the time evolution of the dislocation population 
$N_d$ (Fig.S6, SI\cite{epaps}), a marker that is distinct to polycrystals, 
compared to amorphous systems. 

Distribution of particle displacements $P(u)$, where $u$ is the magnitude of the 2D displacement,
is an useful marker for spatio-temporal heterogeneity \cite{santi,Zapperi}. 
In Fig.\ref{fig2}(d) we show that $P(u)$, collected after different time intervals 
$t$ (see Fig.S6, SI\cite{epaps}), collapse 
onto a master curve, as should be the case for steady state conditions, 
when $P(u)$ and $u$ are appropriately scaled. 
This implies a stationary pattern of displacements originating from different regions of the
channel and the scaling can be understood from the average velocity profile.   
The dominant peak of $P(u)$ in Fig.\ref{fig2}-b is populated by  particles 
from the bulk region that exhibits plug flow. The nearly horizontal part of $P(u)$ 
comes from the boundary layers and can be rationalised as follows. The mean velocity 
profile $v(y)\approx\frac{u}{t}$ is approximately linear in $y$, the distance from the walls.
In the distribution the particles with displacements between $u$ and $u+du$ come 
from the strip of width $dy$ and length $L$, thus $P(u)du \propto dy.L$. The linear profile 
near the wall gives $dy \propto du/t$. Using this we get $P(u)t = $constant, i.e, $P(u)t$ is 
independent of $u$. 
Furthermore, to characterise the velocity fluctuation we define a non-dimensional velocity 
$\Delta v_x=\frac{v_x-\langle v_x (y)\rangle}{v_{rms}}$ where $v_x$ is the instantaneous velocity along $\hat x$, 
$\langle v_x(y)\rangle $ is the average velocity and $v_{rms}$ is the rms velocity. The non-dimensional 
velocity fluctuation shows a non-Gaussian distribution that deviates at the tails, see inset of 
Fig.\ref{fig2}-b. Interestingly, similar non-Gaussian distribution has been reported 
in dense granular flows \cite{nott_expt,orpe,nott_review}. However, the distribution is highly asymmetric
 in case of dense granular flows which is  notably different with polycrystal.

\begin{figure}[]
\centerline{\includegraphics[width=8cm]{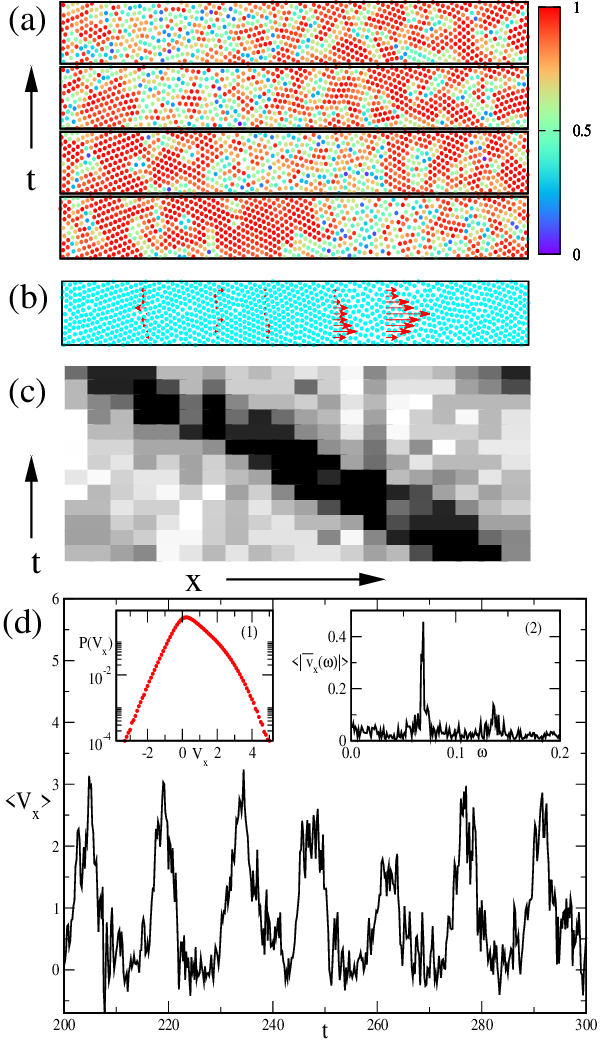}}
 \caption{Fluidization of a flowing polycrystal in narrow channel ($w=9.2,\sigma_w=6.4$). 
(a) The four panels show a moving patch of fluid, with time increasing upwards. 
Colour bar shows value of $\Psi_6$ OP.
(b) Shows heterogeneous velocity map in the earliest snapshot, on five vertical strips.
In the corresponding density map (c), the fluid front moves left, opposite to 
the main flow; darker the colour, lower the density. 
(d) Average particle velocity $\langle v_x\rangle$ in an area element at the 
channel centre. Nearly periodic oscillations imply that the liquid patch 
moves at a constant speed across the channel. 
Inset-1 shows skewed velocity distribution $P(v_x)$ of all particles.
The peak in inset-2 confirms the periodicity in $\langle v_x\rangle$.}
\label{fig3}
\end{figure}

Qualitatively new flow behaviour emerges when the channel is narrow $(w\approx{10d})$,
and the rheological response is probed in the vicinity of the yield threshold ($\sigma_w=6.4$);
see the marked location in Fig.\ref{fig.phase_dia}. 
The moving polycrystal disintegrates into chunks of crystallites and fluid-like 
patches, along the length of the channel; see Fig.\ref{fig3}. In the upper four panels (Fig.\ref{fig3}(a)), 
we show time evolution of
snapshots of  particle positions, with time increasing upwards as indicated, using
the  $\Psi_6$ representation which clearly differentiates the crystalline and fluid zones.
We also observe that the density in the fluid zone is  relatively less. In 
Fig.\ref{fig3}(b),  we  superpose the particle velocities, in few narrow vertical strips, 
corresponding to the panel just above it. 
It is evident that the particles in the fluid zones move fast while the 
crystalline zones are nearly static with small random velocities. Yet, 
the flow profile $v_x(y)$, averaged along the flow  direction  ($\hat x$), 
is parabolic (see inset of Fig.\ref{fig.phase_dia}), which is mainly contributed   
by the liquid like regions, where the profile is strongly parabolic. 
Thus, from the average flow profile, it is not possible to guess about 
the underlying density stratification.  Although the solid-liquid
interfaces are not sharp in Fig.\ref{fig3}, it appears  that the 
interface  is moving left, in the opposite direction of the main flow.
Before probing this in details, we note that 
density waves were reported in gravity-driven granular flows and  traffic flows 
\cite{densitywavegranular,kerner1993cluster}. 
Also, in experiments with driven, dense colloidal suspensions,
backward moving density waves were reported \cite{poonexpt, experiment_fluid_back_flow} 
and modelled \cite{starktheory}. However, note here, that in our case, 
we have no explicit solvent, as was invoked in discussing such observations in
experiments \cite{poonexpt} or for modeling \cite{starktheory}.

To capture the motion of the diffuse liquid-solid interface, we constructed an effective course 
grained, 1-D density field by integrating the particle number density along $\hat y$.  
Fig.\ref{fig3}(c) shows density snapshots at successive time intervals.
The liquid patch (the low density dark region) can be seen moving to the left at nearly 
constant speed (as the kympgraph is linear). To confirm this effect, in 
Fig\ref{fig3}(d), we recorded the average particle velocity 
$\langle v_x (t)\rangle$ in a fixed area element,  at the centre of the channel, as a 
function of time. Almost regular temporal oscillations can be seen. The corresponding 
Fourier transform of $|\langle \overline{v_x}(\omega)\rangle|$ 
shows a dominant peak and a weak 
first harmonic, see inset-2 in Fig\ref{fig3}(d) . 
In inset-1, we present the velocity distribution $P(v_x)$ of
all the particles in the channel.  The distribution is asymmetric and gets higher weightage 
from the positive side, implying that  the average velocity is rightward, although there
are local motions occurring in the opposite direction as well. We checked that 
the positive tail at high $v_x$ is contributed by the fast moving liquid particles and the 
maxima at small positive $v_x$ comes from the crystalline regions.  
The low density liquid patch moves backward essentially by treadmiling. Consider 
a liquid patch in between two solid blocks. The exposed layer from the left block 
melts into liquid due to $F_{ext}$ and thermal forces, and the "atoms" move right 
due to $F_{ext}$.  This way the left front of the liquid moves one atomic spacing 
to the left. At its right front, right moving "atoms" from the liquid, driven by 
$F_{ext}$ add a new layer to the solid block on the right. So the right front of 
the liquid moves left by one unit.    
The melting occurs as $F_{ext}$ overcomes cohesion force of the LJ solid and
the melting rate is constant provided the number densities in both 
the solid and the liquid attain steady values at a particular temperature. 

Note that in this mechanism a given particle periodically resides
in the solid and liquid regions, and moves slow and fast accordingly.
We rationalize the density wave generation along the same lines as
in Ref\cite{starktheory}, with one significant difference. The body force
in our force balance equation (below) is dependent on particle
number density, as in usual for gravity driven flows.
\begin{eqnarray}
0 &=& \rho F_{ext}-\xi(\rho) v +\nu\frac{\partial ^2 v}{\partial x^2} \; ,
\end{eqnarray}
In this 1D equation, $\rho$ is a spatially-varying number density, $v$
is the velocity, $F_{ext}$ is the applied external body force, 
$\xi$ is the friction coefficient and $\nu$ is the effective viscosity. In 
SI-section-V \cite{epaps}, using linear stability analysis, we show
that the density wave occurs only when the force $F_{ext}$ is below 
some threshold value; we also verify this in our simulations. Such a threshold was absent 
in the findings of Ref\cite{starktheory}.

Plug like flows have been well known in amorphous  yield-stress materials 
\cite{roberts2007direct,conrad2008structure, poonexpt,nikoubashman2012flow}.
What is novel for flowing polycrystal is the coexistence of crystallinity 
and flow. Although velocity heterogeneity leads to remodelling of the 
grains, a polycrystalline morphology is still maintained during the 
plug flow. This is because grain remodelling rates are slow compared 
to the average velocity of the plug flow. Even flow transition from plug to parabolic 
profile at decreased channel 
width has been reported  before for concentrated emulsions 
\cite{goyon,mansard2013molecular,bocquet2009kinetic} and  granular systems 
\cite{kamrin2012nonlocal}. While these have been rationalised via non-local 
rheological models \cite{bocquet2009kinetic}, in our system the transition
is accompanied by clear morphological changes of the structure (schematic 
diagrams in Fig-1), and it occurs when the channel width is comparable to
typical grain size,  the relevant correlation length of our system.   
Thus, within the non-local rheological models, these structural changes at the level of
grains need to be incorporated as well.

\section{Summary}
In summary, we have illustrated how the interplay between grain size, 
the additional length scale in soft polycrystals, and the channel width 
result in  contrasting Poiseuille flow  for wide and narrow 
rough channels. At large channel width, the polycrystalline structure 
is retained in the visibly plug like region at the centre, although the 
continuous breakages and coalescences occur due to influence of 
the boundary. This, obviously, has no parallel in polydisperse 
emulsions or granular systems.
However, for narrow channels, when the channel width becomes 
comparable to typical grain size, the polycrystalline structure 
significantly melts during flow and we observe a transverse density 
stratification, with a backward moving low density wave, analogous 
to flowing granular matter \cite{horikawa1996self}, colloids 
\cite{poonexpt,starktheory,haw2004jamming,campbell2010jamming} 
and even traffic flow  in highways \cite{kerner1993cluster}. 
Interestingly, as  the density stratification switches from parallel to 
transverse direction, at intermediate channel widths the boundary 
layer shows unstable, transverse incursions (Fig.S7, SI\cite{epaps}), 
which stabilise as the width is further reduced.
Our work should motivate further experimental exploration of such flow behaviour 
in colloidal polycrystals, in the context of increased interest in such materials.

\bibliography{letter_v5}

\begin{thebibliography}{36}%
\makeatletter
\providecommand \@ifxundefined [1]{%
 \@ifx{#1\undefined}
}%
\providecommand \@ifnum [1]{%
 \ifnum #1\expandafter \@firstoftwo
 \else \expandafter \@secondoftwo
 \fi
}%
\providecommand \@ifx [1]{%
 \ifx #1\expandafter \@firstoftwo
 \else \expandafter \@secondoftwo
 \fi
}%
\providecommand \natexlab [1]{#1}%
\providecommand \enquote  [1]{``#1''}%
\providecommand \bibnamefont  [1]{#1}%
\providecommand \bibfnamefont [1]{#1}%
\providecommand \citenamefont [1]{#1}%
\providecommand \href@noop [0]{\@secondoftwo}%
\providecommand \href [0]{\begingroup \@sanitize@url \@href}%
\providecommand \@href[1]{\@@startlink{#1}\@@href}%
\providecommand \@@href[1]{\endgroup#1\@@endlink}%
\providecommand \@sanitize@url [0]{\catcode `\\12\catcode `\$12\catcode
  `\&12\catcode `\#12\catcode `\^12\catcode `\_12\catcode `\%12\relax}%
\providecommand \@@startlink[1]{}%
\providecommand \@@endlink[0]{}%
\providecommand \url  [0]{\begingroup\@sanitize@url \@url }%
\providecommand \@url [1]{\endgroup\@href {#1}{\urlprefix }}%
\providecommand \urlprefix  [0]{URL }%
\providecommand \Eprint [0]{\href }%
\providecommand \doibase [0]{http://dx.doi.org/}%
\providecommand \selectlanguage [0]{\@gobble}%
\providecommand \bibinfo  [0]{\@secondoftwo}%
\providecommand \bibfield  [0]{\@secondoftwo}%
\providecommand \translation [1]{[#1]}%
\providecommand \BibitemOpen [0]{}%
\providecommand \bibitemStop [0]{}%
\providecommand \bibitemNoStop [0]{.\EOS\space}%
\providecommand \EOS [0]{\spacefactor3000\relax}%
\providecommand \BibitemShut  [1]{\csname bibitem#1\endcsname}%
\let\auto@bib@innerbib\@empty
\bibitem [{\citenamefont {Gokhale}\ \emph {et~al.}(2013)\citenamefont
  {Gokhale}, \citenamefont {Nagamanasa}, \citenamefont {Ganapathy},\ and\
  \citenamefont {Sood}}]{gokhale2013grain}%
  \BibitemOpen
  \bibfield  {author} {\bibinfo {author} {\bibfnamefont {S.}~\bibnamefont
  {Gokhale}}, \bibinfo {author} {\bibfnamefont {K.~H.}\ \bibnamefont
  {Nagamanasa}}, \bibinfo {author} {\bibfnamefont {R.}~\bibnamefont
  {Ganapathy}}, \ and\ \bibinfo {author} {\bibfnamefont {A.}~\bibnamefont
  {Sood}},\ }\href@noop {} {\bibfield  {journal} {\bibinfo  {journal} {Soft
  Matter}\ }\textbf {\bibinfo {volume} {9}},\ \bibinfo {pages} {6634} (\bibinfo
  {year} {2013})}\BibitemShut {NoStop}%
\bibitem [{\citenamefont {Zhang}\ and\ \citenamefont
  {Han}(2018)}]{zhang2018compression}%
  \BibitemOpen
  \bibfield  {author} {\bibinfo {author} {\bibfnamefont {H.}~\bibnamefont
  {Zhang}}\ and\ \bibinfo {author} {\bibfnamefont {Y.}~\bibnamefont {Han}},\
  }\href@noop {} {\bibfield  {journal} {\bibinfo  {journal} {Physical Review
  X}\ }\textbf {\bibinfo {volume} {8}},\ \bibinfo {pages} {041023} (\bibinfo
  {year} {2018})}\BibitemShut {NoStop}%
\bibitem [{\citenamefont {Biswas}\ \emph {et~al.}(2013)\citenamefont {Biswas},
  \citenamefont {Grant}, \citenamefont {Samajdar}, \citenamefont {Haldar},\
  and\ \citenamefont {Sain}}]{santi}%
  \BibitemOpen
  \bibfield  {author} {\bibinfo {author} {\bibfnamefont {S.}~\bibnamefont
  {Biswas}}, \bibinfo {author} {\bibfnamefont {M.}~\bibnamefont {Grant}},
  \bibinfo {author} {\bibfnamefont {I.}~\bibnamefont {Samajdar}}, \bibinfo
  {author} {\bibfnamefont {A.}~\bibnamefont {Haldar}}, \ and\ \bibinfo {author}
  {\bibfnamefont {A.}~\bibnamefont {Sain}},\ }\href@noop {} {\bibfield
  {journal} {\bibinfo  {journal} {Scientific reports}\ }\textbf {\bibinfo
  {volume} {3}} (\bibinfo {year} {2013})}\BibitemShut {NoStop}%
\bibitem [{\citenamefont {Gokhale}\ \emph {et~al.}(2012)\citenamefont
  {Gokhale}, \citenamefont {Nagamanasa}, \citenamefont {Santhosh},
  \citenamefont {Sood},\ and\ \citenamefont
  {Ganapathy}}]{gokhale2012directional}%
  \BibitemOpen
  \bibfield  {author} {\bibinfo {author} {\bibfnamefont {S.}~\bibnamefont
  {Gokhale}}, \bibinfo {author} {\bibfnamefont {K.~H.}\ \bibnamefont
  {Nagamanasa}}, \bibinfo {author} {\bibfnamefont {V.}~\bibnamefont
  {Santhosh}}, \bibinfo {author} {\bibfnamefont {A.}~\bibnamefont {Sood}}, \
  and\ \bibinfo {author} {\bibfnamefont {R.}~\bibnamefont {Ganapathy}},\
  }\href@noop {} {\bibfield  {journal} {\bibinfo  {journal} {Proceedings of the
  National Academy of Sciences}\ }\textbf {\bibinfo {volume} {109}},\ \bibinfo
  {pages} {20314} (\bibinfo {year} {2012})}\BibitemShut {NoStop}%
\bibitem [{\citenamefont {Tamborini}\ \emph {et~al.}(2014)\citenamefont
  {Tamborini}, \citenamefont {Cipelletti},\ and\ \citenamefont
  {Ramos}}]{tamborini2014plasticity}%
  \BibitemOpen
  \bibfield  {author} {\bibinfo {author} {\bibfnamefont {E.}~\bibnamefont
  {Tamborini}}, \bibinfo {author} {\bibfnamefont {L.}~\bibnamefont
  {Cipelletti}}, \ and\ \bibinfo {author} {\bibfnamefont {L.}~\bibnamefont
  {Ramos}},\ }\href@noop {} {\bibfield  {journal} {\bibinfo  {journal}
  {Physical review letters}\ }\textbf {\bibinfo {volume} {113}},\ \bibinfo
  {pages} {078301} (\bibinfo {year} {2014})}\BibitemShut {NoStop}%
\bibitem [{\citenamefont {Buttinoni}\ \emph {et~al.}(2017)\citenamefont
  {Buttinoni}, \citenamefont {Steinacher}, \citenamefont {Spanke},
  \citenamefont {Pokki}, \citenamefont {Bahmann}, \citenamefont {Nelson},
  \citenamefont {Foffi},\ and\ \citenamefont {Isa}}]{buttinoni2017colloidal}%
  \BibitemOpen
  \bibfield  {author} {\bibinfo {author} {\bibfnamefont {I.}~\bibnamefont
  {Buttinoni}}, \bibinfo {author} {\bibfnamefont {M.}~\bibnamefont
  {Steinacher}}, \bibinfo {author} {\bibfnamefont {H.~T.}\ \bibnamefont
  {Spanke}}, \bibinfo {author} {\bibfnamefont {J.}~\bibnamefont {Pokki}},
  \bibinfo {author} {\bibfnamefont {S.}~\bibnamefont {Bahmann}}, \bibinfo
  {author} {\bibfnamefont {B.}~\bibnamefont {Nelson}}, \bibinfo {author}
  {\bibfnamefont {G.}~\bibnamefont {Foffi}}, \ and\ \bibinfo {author}
  {\bibfnamefont {L.}~\bibnamefont {Isa}},\ }\href@noop {} {\bibfield
  {journal} {\bibinfo  {journal} {Physical Review E}\ }\textbf {\bibinfo
  {volume} {95}},\ \bibinfo {pages} {012610} (\bibinfo {year}
  {2017})}\BibitemShut {NoStop}%
\bibitem [{\citenamefont {Lavergne}\ \emph {et~al.}(2018)\citenamefont
  {Lavergne}, \citenamefont {Curran}, \citenamefont {Aarts},\ and\
  \citenamefont {Dullens}}]{Lavergne6922}%
  \BibitemOpen
  \bibfield  {author} {\bibinfo {author} {\bibfnamefont {F.~A.}\ \bibnamefont
  {Lavergne}}, \bibinfo {author} {\bibfnamefont {A.}~\bibnamefont {Curran}},
  \bibinfo {author} {\bibfnamefont {D.~G. A.~L.}\ \bibnamefont {Aarts}}, \ and\
  \bibinfo {author} {\bibfnamefont {R.~P.~A.}\ \bibnamefont {Dullens}},\ }\href
  {\doibase 10.1073/pnas.1804352115} {\bibfield  {journal} {\bibinfo  {journal}
  {Proceedings of the National Academy of Sciences}\ }\textbf {\bibinfo
  {volume} {115}},\ \bibinfo {pages} {6922} (\bibinfo {year}
  {2018})}\BibitemShut {NoStop}%
\bibitem [{\citenamefont {Shiba}\ and\ \citenamefont
  {Onuki}(2010)}]{shiba2010plastic}%
  \BibitemOpen
  \bibfield  {author} {\bibinfo {author} {\bibfnamefont {H.}~\bibnamefont
  {Shiba}}\ and\ \bibinfo {author} {\bibfnamefont {A.}~\bibnamefont {Onuki}},\
  }\href@noop {} {\bibfield  {journal} {\bibinfo  {journal} {Physical Review
  E}\ }\textbf {\bibinfo {volume} {81}},\ \bibinfo {pages} {051501} (\bibinfo
  {year} {2010})}\BibitemShut {NoStop}%
\bibitem [{\citenamefont {Hamanaka}\ and\ \citenamefont {Onuki}(2006)}]{Onuki}%
  \BibitemOpen
  \bibfield  {author} {\bibinfo {author} {\bibfnamefont {T.}~\bibnamefont
  {Hamanaka}}\ and\ \bibinfo {author} {\bibfnamefont {A.}~\bibnamefont
  {Onuki}},\ }\href@noop {} {\bibfield  {journal} {\bibinfo  {journal}
  {Physical Review E}\ }\textbf {\bibinfo {volume} {74}},\ \bibinfo {pages}
  {011506} (\bibinfo {year} {2006})}\BibitemShut {NoStop}%
\bibitem [{\citenamefont {Sarkar}\ \emph {et~al.}(2017)\citenamefont {Sarkar},
  \citenamefont {Biswas}, \citenamefont {Chaudhuri},\ and\ \citenamefont
  {Sain}}]{sarkar2017grain}%
  \BibitemOpen
  \bibfield  {author} {\bibinfo {author} {\bibfnamefont {T.}~\bibnamefont
  {Sarkar}}, \bibinfo {author} {\bibfnamefont {S.}~\bibnamefont {Biswas}},
  \bibinfo {author} {\bibfnamefont {P.}~\bibnamefont {Chaudhuri}}, \ and\
  \bibinfo {author} {\bibfnamefont {A.}~\bibnamefont {Sain}},\ }\href@noop {}
  {\bibfield  {journal} {\bibinfo  {journal} {Physical Review Materials}\
  }\textbf {\bibinfo {volume} {1}},\ \bibinfo {pages} {070601} (\bibinfo {year}
  {2017})}\BibitemShut {NoStop}%
\bibitem [{\citenamefont {Goyon}\ \emph {et~al.}(2008)\citenamefont {Goyon},
  \citenamefont {Colin}, \citenamefont {Ovarlez}, \citenamefont {Ajdari},\ and\
  \citenamefont {Bocquet}}]{goyon}%
  \BibitemOpen
  \bibfield  {author} {\bibinfo {author} {\bibfnamefont {J.}~\bibnamefont
  {Goyon}}, \bibinfo {author} {\bibfnamefont {A.}~\bibnamefont {Colin}},
  \bibinfo {author} {\bibfnamefont {G.}~\bibnamefont {Ovarlez}}, \bibinfo
  {author} {\bibfnamefont {A.}~\bibnamefont {Ajdari}}, \ and\ \bibinfo {author}
  {\bibfnamefont {L.}~\bibnamefont {Bocquet}},\ }\href@noop {} {\bibfield
  {journal} {\bibinfo  {journal} {Nature}\ }\textbf {\bibinfo {volume} {454}},\
  \bibinfo {pages} {84} (\bibinfo {year} {2008})}\BibitemShut {NoStop}%
\bibitem [{\citenamefont {Mansard}\ \emph {et~al.}(2013)\citenamefont
  {Mansard}, \citenamefont {Colin}, \citenamefont {Chaudhuri},\ and\
  \citenamefont {Bocquet}}]{mansard2013molecular}%
  \BibitemOpen
  \bibfield  {author} {\bibinfo {author} {\bibfnamefont {V.}~\bibnamefont
  {Mansard}}, \bibinfo {author} {\bibfnamefont {A.}~\bibnamefont {Colin}},
  \bibinfo {author} {\bibfnamefont {P.}~\bibnamefont {Chaudhuri}}, \ and\
  \bibinfo {author} {\bibfnamefont {L.}~\bibnamefont {Bocquet}},\ }\href@noop
  {} {\bibfield  {journal} {\bibinfo  {journal} {Soft matter}\ }\textbf
  {\bibinfo {volume} {9}},\ \bibinfo {pages} {7489} (\bibinfo {year}
  {2013})}\BibitemShut {NoStop}%
\bibitem [{\citenamefont {Chaudhuri}\ and\ \citenamefont
  {Horbach}(2014)}]{chaudhuri2014poiseuille}%
  \BibitemOpen
  \bibfield  {author} {\bibinfo {author} {\bibfnamefont {P.}~\bibnamefont
  {Chaudhuri}}\ and\ \bibinfo {author} {\bibfnamefont {J.}~\bibnamefont
  {Horbach}},\ }\href@noop {} {\bibfield  {journal} {\bibinfo  {journal}
  {Physical Review E}\ }\textbf {\bibinfo {volume} {90}},\ \bibinfo {pages}
  {040301} (\bibinfo {year} {2014})}\BibitemShut {NoStop}%
\bibitem [{\citenamefont {Chaudhuri}\ \emph {et~al.}(2012)\citenamefont
  {Chaudhuri}, \citenamefont {Mansard}, \citenamefont {Colin},\ and\
  \citenamefont {Bocquet}}]{chaudhuri2012dynamical}%
  \BibitemOpen
  \bibfield  {author} {\bibinfo {author} {\bibfnamefont {P.}~\bibnamefont
  {Chaudhuri}}, \bibinfo {author} {\bibfnamefont {V.}~\bibnamefont {Mansard}},
  \bibinfo {author} {\bibfnamefont {A.}~\bibnamefont {Colin}}, \ and\ \bibinfo
  {author} {\bibfnamefont {L.}~\bibnamefont {Bocquet}},\ }\href@noop {}
  {\bibfield  {journal} {\bibinfo  {journal} {Physical review letters}\
  }\textbf {\bibinfo {volume} {109}},\ \bibinfo {pages} {036001} (\bibinfo
  {year} {2012})}\BibitemShut {NoStop}%
\bibitem [{epa(ctra)}]{epaps}%
  \BibitemOpen
  \href@noop {} {\bibfield  {journal} {\bibinfo  {journal} {Supplementary
  Material (with more analysis), which includes Refs.}\ } (\bibinfo {year}
  {\cite{starktheory,daub2009stick,Goldenfeld_power_spectra}})}\BibitemShut
  {NoStop}%
\bibitem [{\citenamefont {Varnik}\ and\ \citenamefont
  {Raabe}(2008)}]{varnik2008profile}%
  \BibitemOpen
  \bibfield  {author} {\bibinfo {author} {\bibfnamefont {F.}~\bibnamefont
  {Varnik}}\ and\ \bibinfo {author} {\bibfnamefont {D.}~\bibnamefont {Raabe}},\
  }\href@noop {} {\bibfield  {journal} {\bibinfo  {journal} {Physical Review
  E}\ }\textbf {\bibinfo {volume} {77}},\ \bibinfo {pages} {011504} (\bibinfo
  {year} {2008})}\BibitemShut {NoStop}%
\bibitem [{\citenamefont {Genovese}\ and\ \citenamefont
  {Sprakel}(2011)}]{genovese2011crystallization}%
  \BibitemOpen
  \bibfield  {author} {\bibinfo {author} {\bibfnamefont {D.}~\bibnamefont
  {Genovese}}\ and\ \bibinfo {author} {\bibfnamefont {J.}~\bibnamefont
  {Sprakel}},\ }\href@noop {} {\bibfield  {journal} {\bibinfo  {journal} {Soft
  Matter}\ }\textbf {\bibinfo {volume} {7}},\ \bibinfo {pages} {3889} (\bibinfo
  {year} {2011})}\BibitemShut {NoStop}%
\bibitem [{\citenamefont {Jana}\ \emph {et~al.}(2017)\citenamefont {Jana},
  \citenamefont {Alava},\ and\ \citenamefont {Zapperi}}]{Zapperi}%
  \BibitemOpen
  \bibfield  {author} {\bibinfo {author} {\bibfnamefont {P.~K.}\ \bibnamefont
  {Jana}}, \bibinfo {author} {\bibfnamefont {M.~J.}\ \bibnamefont {Alava}}, \
  and\ \bibinfo {author} {\bibfnamefont {S.}~\bibnamefont {Zapperi}},\
  }\href@noop {} {\bibfield  {journal} {\bibinfo  {journal} {Scientific
  Reports}\ }\textbf {\bibinfo {volume} {7}},\ \bibinfo {pages} {45550}
  (\bibinfo {year} {2017})}\BibitemShut {NoStop}%
\bibitem [{\citenamefont {Moka}\ and\ \citenamefont {Nott}(2005)}]{nott_expt}%
  \BibitemOpen
  \bibfield  {author} {\bibinfo {author} {\bibfnamefont {S.}~\bibnamefont
  {Moka}}\ and\ \bibinfo {author} {\bibfnamefont {P.~R.}\ \bibnamefont
  {Nott}},\ }\href@noop {} {\bibfield  {journal} {\bibinfo  {journal} {Physical
  review letters}\ }\textbf {\bibinfo {volume} {95}},\ \bibinfo {pages}
  {068003} (\bibinfo {year} {2005})}\BibitemShut {NoStop}%
\bibitem [{\citenamefont {Orpe}\ and\ \citenamefont {Kudrolli}(2007)}]{orpe}%
  \BibitemOpen
  \bibfield  {author} {\bibinfo {author} {\bibfnamefont {A.~V.}\ \bibnamefont
  {Orpe}}\ and\ \bibinfo {author} {\bibfnamefont {A.}~\bibnamefont
  {Kudrolli}},\ }\href@noop {} {\bibfield  {journal} {\bibinfo  {journal}
  {Physical review letters}\ }\textbf {\bibinfo {volume} {98}},\ \bibinfo
  {pages} {238001} (\bibinfo {year} {2007})}\BibitemShut {NoStop}%
\bibitem [{\citenamefont {Ananda}\ \emph {et~al.}(2008)\citenamefont {Ananda},
  \citenamefont {Moka},\ and\ \citenamefont {Nott}}]{nott_review}%
  \BibitemOpen
  \bibfield  {author} {\bibinfo {author} {\bibfnamefont {K.}~\bibnamefont
  {Ananda}}, \bibinfo {author} {\bibfnamefont {S.}~\bibnamefont {Moka}}, \ and\
  \bibinfo {author} {\bibfnamefont {P.~R.}\ \bibnamefont {Nott}},\ }\href@noop
  {} {\bibfield  {journal} {\bibinfo  {journal} {Journal of Fluid Mechanics}\
  }\textbf {\bibinfo {volume} {610}},\ \bibinfo {pages} {69} (\bibinfo {year}
  {2008})}\BibitemShut {NoStop}%
\bibitem [{\citenamefont {Liss}\ \emph {et~al.}(2002)\citenamefont {Liss},
  \citenamefont {Conway},\ and\ \citenamefont {Glasser}}]{densitywavegranular}%
  \BibitemOpen
  \bibfield  {author} {\bibinfo {author} {\bibfnamefont {E.~D.}\ \bibnamefont
  {Liss}}, \bibinfo {author} {\bibfnamefont {S.~L.}\ \bibnamefont {Conway}}, \
  and\ \bibinfo {author} {\bibfnamefont {B.~J.}\ \bibnamefont {Glasser}},\
  }\href@noop {} {\bibfield  {journal} {\bibinfo  {journal} {Physics of
  Fluids}\ }\textbf {\bibinfo {volume} {14}},\ \bibinfo {pages} {3309}
  (\bibinfo {year} {2002})}\BibitemShut {NoStop}%
\bibitem [{\citenamefont {Kerner}\ and\ \citenamefont
  {Konh{\"a}user}(1993)}]{kerner1993cluster}%
  \BibitemOpen
  \bibfield  {author} {\bibinfo {author} {\bibfnamefont {B.~S.}\ \bibnamefont
  {Kerner}}\ and\ \bibinfo {author} {\bibfnamefont {P.}~\bibnamefont
  {Konh{\"a}user}},\ }\href@noop {} {\bibfield  {journal} {\bibinfo  {journal}
  {Physical Review E}\ }\textbf {\bibinfo {volume} {48}},\ \bibinfo {pages}
  {R2335} (\bibinfo {year} {1993})}\BibitemShut {NoStop}%
\bibitem [{\citenamefont {Isa}\ \emph {et~al.}(2009)\citenamefont {Isa},
  \citenamefont {Besseling}, \citenamefont {Morozov},\ and\ \citenamefont
  {Poon}}]{poonexpt}%
  \BibitemOpen
  \bibfield  {author} {\bibinfo {author} {\bibfnamefont {L.}~\bibnamefont
  {Isa}}, \bibinfo {author} {\bibfnamefont {R.}~\bibnamefont {Besseling}},
  \bibinfo {author} {\bibfnamefont {A.~N.}\ \bibnamefont {Morozov}}, \ and\
  \bibinfo {author} {\bibfnamefont {W.~C.~K.}\ \bibnamefont {Poon}},\ }\href
  {\doibase 10.1103/PhysRevLett.102.058302} {\bibfield  {journal} {\bibinfo
  {journal} {Phys. Rev. Lett.}\ }\textbf {\bibinfo {volume} {102}},\ \bibinfo
  {pages} {058302} (\bibinfo {year} {2009})}\BibitemShut {NoStop}%
\bibitem [{\citenamefont {Horikawa}\ \emph {et~al.}(1995)\citenamefont
  {Horikawa}, \citenamefont {Nakahara}, \citenamefont {Nakayama},\ and\
  \citenamefont {Matsushita}}]{experiment_fluid_back_flow}%
  \BibitemOpen
  \bibfield  {author} {\bibinfo {author} {\bibfnamefont {S.}~\bibnamefont
  {Horikawa}}, \bibinfo {author} {\bibfnamefont {A.}~\bibnamefont {Nakahara}},
  \bibinfo {author} {\bibfnamefont {T.}~\bibnamefont {Nakayama}}, \ and\
  \bibinfo {author} {\bibfnamefont {M.}~\bibnamefont {Matsushita}},\
  }\href@noop {} {\bibfield  {journal} {\bibinfo  {journal} {Journal of the
  Physical Society of Japan}\ }\textbf {\bibinfo {volume} {64}},\ \bibinfo
  {pages} {1870} (\bibinfo {year} {1995})}\BibitemShut {NoStop}%
\bibitem [{\citenamefont {Kanehl}\ and\ \citenamefont
  {Stark}(2017)}]{starktheory}%
  \BibitemOpen
  \bibfield  {author} {\bibinfo {author} {\bibfnamefont {P.}~\bibnamefont
  {Kanehl}}\ and\ \bibinfo {author} {\bibfnamefont {H.}~\bibnamefont {Stark}},\
  }\href {\doibase 10.1103/PhysRevLett.119.018002} {\bibfield  {journal}
  {\bibinfo  {journal} {Phys. Rev. Lett.}\ }\textbf {\bibinfo {volume} {119}},\
  \bibinfo {pages} {018002} (\bibinfo {year} {2017})}\BibitemShut {NoStop}%
\bibitem [{\citenamefont {Roberts}\ \emph {et~al.}(2007)\citenamefont
  {Roberts}, \citenamefont {Mohraz}, \citenamefont {Christensen},\ and\
  \citenamefont {Lewis}}]{roberts2007direct}%
  \BibitemOpen
  \bibfield  {author} {\bibinfo {author} {\bibfnamefont {M.~T.}\ \bibnamefont
  {Roberts}}, \bibinfo {author} {\bibfnamefont {A.}~\bibnamefont {Mohraz}},
  \bibinfo {author} {\bibfnamefont {K.~T.}\ \bibnamefont {Christensen}}, \ and\
  \bibinfo {author} {\bibfnamefont {J.~A.}\ \bibnamefont {Lewis}},\ }\href@noop
  {} {\bibfield  {journal} {\bibinfo  {journal} {Langmuir}\ }\textbf {\bibinfo
  {volume} {23}},\ \bibinfo {pages} {8726} (\bibinfo {year}
  {2007})}\BibitemShut {NoStop}%
\bibitem [{\citenamefont {Conrad}\ and\ \citenamefont
  {Lewis}(2008)}]{conrad2008structure}%
  \BibitemOpen
  \bibfield  {author} {\bibinfo {author} {\bibfnamefont {J.~C.}\ \bibnamefont
  {Conrad}}\ and\ \bibinfo {author} {\bibfnamefont {J.~A.}\ \bibnamefont
  {Lewis}},\ }\href@noop {} {\bibfield  {journal} {\bibinfo  {journal}
  {Langmuir}\ }\textbf {\bibinfo {volume} {24}},\ \bibinfo {pages} {7628}
  (\bibinfo {year} {2008})}\BibitemShut {NoStop}%
\bibitem [{\citenamefont {Nikoubashman}\ \emph {et~al.}(2012)\citenamefont
  {Nikoubashman}, \citenamefont {Kahl},\ and\ \citenamefont
  {Likos}}]{nikoubashman2012flow}%
  \BibitemOpen
  \bibfield  {author} {\bibinfo {author} {\bibfnamefont {A.}~\bibnamefont
  {Nikoubashman}}, \bibinfo {author} {\bibfnamefont {G.}~\bibnamefont {Kahl}},
  \ and\ \bibinfo {author} {\bibfnamefont {C.~N.}\ \bibnamefont {Likos}},\
  }\href@noop {} {\bibfield  {journal} {\bibinfo  {journal} {Soft Matter}\
  }\textbf {\bibinfo {volume} {8}},\ \bibinfo {pages} {4121} (\bibinfo {year}
  {2012})}\BibitemShut {NoStop}%
\bibitem [{\citenamefont {Bocquet}\ \emph {et~al.}(2009)\citenamefont
  {Bocquet}, \citenamefont {Colin},\ and\ \citenamefont
  {Ajdari}}]{bocquet2009kinetic}%
  \BibitemOpen
  \bibfield  {author} {\bibinfo {author} {\bibfnamefont {L.}~\bibnamefont
  {Bocquet}}, \bibinfo {author} {\bibfnamefont {A.}~\bibnamefont {Colin}}, \
  and\ \bibinfo {author} {\bibfnamefont {A.}~\bibnamefont {Ajdari}},\
  }\href@noop {} {\bibfield  {journal} {\bibinfo  {journal} {Physical review
  letters}\ }\textbf {\bibinfo {volume} {103}},\ \bibinfo {pages} {036001}
  (\bibinfo {year} {2009})}\BibitemShut {NoStop}%
\bibitem [{\citenamefont {Kamrin}\ and\ \citenamefont
  {Koval}(2012)}]{kamrin2012nonlocal}%
  \BibitemOpen
  \bibfield  {author} {\bibinfo {author} {\bibfnamefont {K.}~\bibnamefont
  {Kamrin}}\ and\ \bibinfo {author} {\bibfnamefont {G.}~\bibnamefont {Koval}},\
  }\href@noop {} {\bibfield  {journal} {\bibinfo  {journal} {Physical Review
  Letters}\ }\textbf {\bibinfo {volume} {108}},\ \bibinfo {pages} {178301}
  (\bibinfo {year} {2012})}\BibitemShut {NoStop}%
\bibitem [{\citenamefont {Horikawa}\ \emph {et~al.}(1996)\citenamefont
  {Horikawa}, \citenamefont {Isoda}, \citenamefont {Nakayama}, \citenamefont
  {Nakahara},\ and\ \citenamefont {Matsushita}}]{horikawa1996self}%
  \BibitemOpen
  \bibfield  {author} {\bibinfo {author} {\bibfnamefont {S.}~\bibnamefont
  {Horikawa}}, \bibinfo {author} {\bibfnamefont {T.}~\bibnamefont {Isoda}},
  \bibinfo {author} {\bibfnamefont {T.}~\bibnamefont {Nakayama}}, \bibinfo
  {author} {\bibfnamefont {A.}~\bibnamefont {Nakahara}}, \ and\ \bibinfo
  {author} {\bibfnamefont {M.}~\bibnamefont {Matsushita}},\ }\href@noop {}
  {\bibfield  {journal} {\bibinfo  {journal} {Physica A: Statistical Mechanics
  and its Applications}\ }\textbf {\bibinfo {volume} {233}},\ \bibinfo {pages}
  {699} (\bibinfo {year} {1996})}\BibitemShut {NoStop}%
\bibitem [{\citenamefont {Haw}(2004)}]{haw2004jamming}%
  \BibitemOpen
  \bibfield  {author} {\bibinfo {author} {\bibfnamefont {M.}~\bibnamefont
  {Haw}},\ }\href@noop {} {\bibfield  {journal} {\bibinfo  {journal} {Physical
  review letters}\ }\textbf {\bibinfo {volume} {92}},\ \bibinfo {pages}
  {185506} (\bibinfo {year} {2004})}\BibitemShut {NoStop}%
\bibitem [{\citenamefont {Campbell}\ and\ \citenamefont
  {Haw}(2010)}]{campbell2010jamming}%
  \BibitemOpen
  \bibfield  {author} {\bibinfo {author} {\bibfnamefont {A.~I.}\ \bibnamefont
  {Campbell}}\ and\ \bibinfo {author} {\bibfnamefont {M.~D.}\ \bibnamefont
  {Haw}},\ }\href@noop {} {\bibfield  {journal} {\bibinfo  {journal} {Soft
  Matter}\ }\textbf {\bibinfo {volume} {6}},\ \bibinfo {pages} {4688} (\bibinfo
  {year} {2010})}\BibitemShut {NoStop}%
\bibitem [{\citenamefont {Daub}\ and\ \citenamefont
  {Carlson}(2009)}]{daub2009stick}%
  \BibitemOpen
  \bibfield  {author} {\bibinfo {author} {\bibfnamefont {E.~G.}\ \bibnamefont
  {Daub}}\ and\ \bibinfo {author} {\bibfnamefont {J.~M.}\ \bibnamefont
  {Carlson}},\ }\href@noop {} {\bibfield  {journal} {\bibinfo  {journal}
  {Physical Review E}\ }\textbf {\bibinfo {volume} {80}},\ \bibinfo {pages}
  {066113} (\bibinfo {year} {2009})}\BibitemShut {NoStop}%
\bibitem [{\citenamefont {Tarp}\ \emph {et~al.}(2014)\citenamefont {Tarp},
  \citenamefont {Angheluta}, \citenamefont {Mathiesen},\ and\ \citenamefont
  {Goldenfeld}}]{Goldenfeld_power_spectra}%
  \BibitemOpen
  \bibfield  {author} {\bibinfo {author} {\bibfnamefont {J.~M.}\ \bibnamefont
  {Tarp}}, \bibinfo {author} {\bibfnamefont {L.}~\bibnamefont {Angheluta}},
  \bibinfo {author} {\bibfnamefont {J.}~\bibnamefont {Mathiesen}}, \ and\
  \bibinfo {author} {\bibfnamefont {N.}~\bibnamefont {Goldenfeld}},\
  }\href@noop {} {\bibfield  {journal} {\bibinfo  {journal} {Physical review
  letters}\ }\textbf {\bibinfo {volume} {113}},\ \bibinfo {pages} {265503}
  (\bibinfo {year} {2014})}\BibitemShut {NoStop}%
\end{thebibliography}%

\clearpage
\newpage

\end{document}